\documentclass[letterpaper]{article} 
\usepackage{aaai25}  
\usepackage{times}  
\usepackage{helvet}  
\usepackage{courier}  
\usepackage[hyphens]{url}  
\usepackage{graphicx} 
\usepackage{natbib}  
\usepackage{caption} 
\usepackage{algorithm}
\usepackage{algorithmic}
\usepackage{fontawesome}
\usepackage{colortbl}

\usepackage{newfloat}
\usepackage{listings}
\DeclareCaptionStyle{ruled}{labelfont=normalfont,labelsep=colon,strut=off} 
\floatstyle{ruled}
\newfloat{listing}{tb}{lst}{}
\floatname{listing}{Listing}
\title{The Stories We Govern By: AI, Risk, and the Power of Imaginaries}

\author{
    Ninell Oldenburg\textsuperscript{\rm 1}, 
    Gleb Papyshev\textsuperscript{\rm 2}
    }
\affiliations{
    \textsuperscript{\rm 1} University of Copenhagen, Center for Philosophy of Artificial Intelligence \\
    \textsuperscript{\rm 2} Lingnan University, Department of Government and International Relations, Division of Artificial Intelligence \\
    niol@hum.ku.dk, glebpapyshev@ln.edu.hk
}

\begin{document}

\maketitle

\begin{abstract}
This paper examines how competing sociotechnical imaginaries of artificial intelligence (AI) risk shape governance decisions and regulatory constraints. Drawing on concepts from science and technology studies, we analyse three dominant narrative groups: existential risk proponents, who emphasise catastrophic AGI scenarios; accelerationists, who portray AI as a transformative force to be unleashed; and critical AI scholars, who foreground present-day harms rooted in systemic inequality. Through an analysis of representative manifesto-style texts, we explore how these imaginaries differ across four dimensions: normative visions of the future, diagnoses of the present social order, views on science and technology, and perceived human agency in managing AI risks. Our findings reveal how these narratives embed distinct assumptions about risk and have the potential to progress into policy-making processes by narrowing the space for alternative governance approaches. We argue against speculative dogmatism and for moving beyond deterministic imaginaries toward regulatory strategies that are grounded in pragmatism.
\end{abstract}



\section{Introduction}\label{sec:introduction}

Discussions around the risks of artificial intelligence (AI) are shaped by narratives that define how we envision the role of technology in society and the kinds of futures we prioritise \citep{jasanoff2015future,campolo2020enchanted, witschas2025prefabricated}. \citet{jasanoff2015future} has defined this as \textit{sociotechnical imaginaries} -- rhetorical frames that actively shape institutional agendas, funding priorities, and regulatory pathways by embedding particular visions of risk, responsibility, and progress into governance structures.

Building on the concept of sociotechnical imaginaries, this paper analyses how AI risk narratives constrain government decisions and actions. We thereby contribute to a growing body of work that examines how collective visions of AI influence its development, use, and regulation, and brings to light the values and viewpoints of the groups from which these imaginaries originate \citep{jasanoff2015future, losch2019visionary, bareis2022talking, bakiner2023pluralistic, natale2019if, cave2020introduction}. Important work has been done to look at the representations of AI in the newspapers \citep{johnson2017ai} or policy proposals in different countries \citep{bareis2022talking}, its role in the imagination of utopian or dystopian futures \citep{cave2020introduction}, in perpetuating the existing power structures \citep{campolo2020enchanted}, or in medical care \citep{gao2020public}. We now add a piece by focusing on how sociotechnical imaginaries of AI \textit{risk} influence governance priorities, while the term itself was always closely tied to issues of risk and governance in other areas.

For this, we examine three exemplary manifesto-type works, one for each of three dominant imaginary groups: 1) AI existential risk (X-risk) proponents, who warn of an artificial general intelligence (AGI)-driven apocalypse and push for technical alignment research and global containment strategies; 2) Accelerationists, who frame AI as a civilizational leap whose transformative potential must be unleashed, arguing that regulation is not only premature but a threat to innovation and geopolitical advantage; and 3) Critical AI scholars, who reject these speculative futures and instead foreground the present harms of AI such as surveillance, labor exploitation, and racial bias. They argue that the most urgent risks are already here and deeply entangled with power and inequality. 

We analyse them by zooming in on four different dimensions, namely i) the imaginaries' normative visions of the future, ii) how they see the current social order and values AI's development is embedded in, iii) the broader role of science and technology in AI's development, and lastly iv) the amount of human agency left they see to act on the foreseen risks. By doing so, we aim to create a detailed picture of the assumptions underlying certain narratives and a holistic picture of their visions of the future.

The paper is structured as follows. First, we introduce the reader to the existing literature on politics and technology, the concept of sociotechnical imaginaries \citep{jasanoff2015future}, and the current narratives prevailing in the three target groups. Next, we describe why and how we picked the target groups and their exemplary statements, as well as the categories we used to analyse the sociotechnical imaginaries. Then, we present the results of our analysis alongside the implications this has on governance. Finally, we give a brief overview of existing AI governance approaches and discuss alternative approaches that move beyond deterministic imaginaries, i.e., a pragmatic governance approach.


\section{Background and Related Work}\label{sec:background}

New technological and scientific developments are often embedded within modernist narratives of progress \citep{bareis2022talking}, and examining those thus provides insight into collective aspirations and visions for the future. \citet{jasanoff2015future} calls these collections of expectations \textit{sociotechnical imaginaries}, or collective visions of technology’s role in society. She defines them as 

\begin{quote}
    Collectively held, institutionally stabilized, and publicly performed visions of desirable futures, animated by shared understandings of forms of social life and social order attainable through, and supportive of, advances in science and technology. \citep{jasanoff2015future}
\end{quote}

As ``legal disputes are in their very nature moments of contestation between disparate understandings of the good," \citep{jasanoff2015future} these imaginaries do not just \textit{predict} a technology's trajectory but also \textit{define} political possibilities through emphasising potential benefits, limitations, and transformative capabilities \citep{bareis2022talking}, as we will see below. \citet{beck2021governance} promote imaginaries therefore as a useful analytical tool ``to explore underlying normative, but often inexplicit, rationales and justifications of policy choices for governing emerging technologies”.

Building on these theoretical foundations, scholars have examined algorithms \citep{bucher2019algorithmic} and AI \citep{sartori2023minding,richter2023imaginaries} through the lens of sociotechnical imaginaries and their influences on policy and governance \citep{bareis2022talking,bakiner2023pluralistic}. For example, \citet{bucher2019algorithmic} examines how visions of the mechanisms and goals of a recommendation algorithm (e.g., if it ``knows" something about a person's current life situation and displays respective ads) influence people's perception and experience of this very algorithm (i.e., how it makes them feel). She develops the term ``algorithmic imaginaries" and argues that these play a major role in developing the algorithm itself and are pivotal for understanding their social power.

\citet{bakiner2023pluralistic} analyses 302 response papers to the draft of the EU AI Act, revealing competing imaginaries. These imaginaries concern (1) the fundamental nature of AI and its relationship with society, politics, and law; (2) the extent to which law can or should guide technological advancement; and (3) the appropriate level of legal intervention in scientific and technological debates. In particular, he finds that different stakeholder groups frame AI-related concerns in distinct ways, which reflect pluralistic and often conflicting imaginaries. Moreover, stakeholders disagree with lawmakers on how AI’s characteristics should inform regulation and the ideal form of legal oversight \citep{bakiner2023pluralistic}.

\citet{bareis2022talking} investigate imaginaries of AI in policy documents from China, the U.S., France, and Germany. They find that all four countries construct AI as an inevitable and disruptive force, usually building on narratives of AI as revolutionary technology, the embeddedness in historical technological progress, and the positioning of AI as a core demand for society. In all these narratives, the uncertainty around AI’s future necessitates leadership and policy interventions. The specific imaginaries of the countries, however, diverge in cultural, political, and economic dimensions and, as a result, produce distinct developmental pathways. \citet{bareis2022talking} argue that these imaginaries create a self-reinforcing ``lock-in" effect where investments require returns, and promises demand fulfillment.

Such narratives of the ``unavoidable" have existed since AI’s early days. In 1961, John McCarthy argued that machines would eventually solve the most complex problems, downplaying distinctions between human and machine cognition \citep{greenberger1962management}. Some years later, \citet{good1966speculations} introduced the idea of an ``intelligence explosion," where AI could recursively improve itself. This idea was further popularised by \citet{vinge1993coming} and \citet{kurzweil2005singularity}, who claimed AI’s surpassing of human intelligence was not just possible but inevitable, leading to the Singularity Thesis.

Today, the Singularity Thesis underpins many narratives like the existential risk (X-risk) ones. Figures like \citet{bostrom2014superintelligence} and \citet{tegmark2018life} argue that AGI is an unavoidable milestone, potentially leading to catastrophic outcomes. They advocate for expert-led AI governance, sidelining broader democratic deliberation. These views are influential in the Effective Altruism movement, where AI ``doomsday” predictions shape funding and policy priorities \citep{witschas2025prefabricated}. This deterministic framing shifts governance focus toward speculative long-term risks rather than present-day interventions.

\definecolor{lightgray}{gray}{0.95}

\renewcommand{\arraystretch}{1.3}
\begin{table*}[!ht]
    \centering
    \rowcolors{2}{lightgray}{white}
    \begin{tabular}{m{5.5cm}|m{11.5cm}}
        \textbf{Category}  & \textbf{Questions} \\ \hline \hline 
        
        \faList \hspace{0.2pt} \textbf{Normative Vision of the Future} & What does the society of the future look like according to the organization? Is AI portrayed as a solution to societal problems, a threat, or an enabler of progress? \\ \hline 
        
        \faUniversity \hspace{0.2pt} \textbf{Social Order and Values} & How do the texts describe the role of different social groups in shaping or being affected by technology? Who are the ``winners" and ``losers" in these future visions? What values (e.g., efficiency, equity, control) are emphasised? \\ \hline 

        \faGears \hspace{0.2pt} \textbf{Role of Science and Technology} & How does AI fit into the larger context of technological and scientific progress? Does the narrative position AI as part of a historical progression, or does it present it as a disruptive break? \\ \hline 
       
        \faWrench \hspace{0.2pt} \textbf{Determinism vs. Agency} & Does the text depict AI’s development as inevitable and deterministic, or is there space for human agency in shaping its trajectory? For example, does it stress leadership, innovation, or regulation to shape AI’s direction? \\ 

    \end{tabular}
    \caption{Operationalization of our Analysis}
    \label{tab:operationalization}
\end{table*}

In contrast, Accelerationists see AI as a transformative force for good, claiming it will solve global challenges like inequality and climate change \citep{soufi2024accelerate, maccoll2023cult}. Rooted in Silicon Valley’s libertarian ethos, accelerationists, including figures with ties to US deregulation advocates, argue \textit{against} AI governance \citep{stokes2024valley}. This framing demands various kinds of deregulation to give AI the full power of development as soon as possible.

A third perspective, critical AI scholarship, rejects grand futurist narratives and focuses on present harms such as bias, surveillance, and labor exploitation \citep{gebru2024tescreal}. While opposing AI future determinism, this line of thought reinforces a different fatalism, suggesting AI’s harms are so deeply embedded that governance is futile \citep{benjamin2019captivating, noble2018algorithms}. This perspective is pro-regulation and has informed legal frameworks such as the European AI Act, which focuses on mitigating present harms rather than speculative future risks.

Each of these narratives presents the technology’s trajectory as predetermined, narrowing policy discussions to extreme scenarios rather than advocating for democratic and agile strategies for shaping AI’s role in society. AI is already being deployed in critical areas like healthcare, policing, and finance under little democratic oversight as if its integration was inevitable rather than a political choice \citep{campolo2020enchanted}. Below, we will detail how reinforcing AI as an inevitable force rather than a contested space of intervention, these imaginaries shape policy debates and narrow the range of governance options considered viable: X-risk advocates push for centralised control, accelerationists reject regulation, and some critical scholars see reform as futile.


\section{Methodology}\label{sec:methods}

\renewcommand{\arraystretch}{1.4}
\begin{table}[b!] \centering
    \rowcolors{2}{lightgray}{white}
    \begin{tabular}{m{2.5cm}|m{5cm}}
    \textbf{Imaginary} & \textbf{Source} \\ \hline \hline
       \textbf{Existential Risk} & Machine Intelligence Research Institute (2025). \textit{The Problem}. \\ \hline
       \textbf{Accelerationism} & Marc Andreessen (2023). \textit{The techno-optimist Manifesto}. \\ \hline
       \textbf{Critical AI} & Distributed AI Research Institute (2022). \textit{DAIR Research Philosophy}. \\
    \end{tabular}
    \caption{Investigated texts used for the comparative analysis}
    \label{tab:papers}
\end{table}

We employ a discourse analysis approach to examine how AI risk imaginaries influence governance debates. We operationalised sociotechnical imaginaries into concrete analytical elements (Table~\ref{tab:operationalization}), to focus on four key dimensions that we think contribute to a holistic picture of each of the imaginaries: visions of society's future, political orientations, views on science and technology, and perceived agency levels.

We investigate how the three groups introduced above (X-risk proponents, accelerationists, and critical AI scholars) construct AI's future as inevitable and how these framings shape policy discussions. We chose this classification based on each group's distinct risk assessment, as this serves as the most direct criterion for differentiating AI risk governance narratives. While these categories intersect with other dimensions (political views, professional backgrounds, demographics, geography) and knowledge groups (activists, academics, indigenous), we leave the development of a comprehensive taxonomy for future research, as it falls outside this paper's scope.

We conducted textual analysis on one foundational text from each group, selecting organizations that publish manifesto-style research agendas and hold prominent positions within their communities. We analysed MIRI (the largest and most established X-risk organization), Marc Andreessen (representing accelerationism, where formal organizations are less common), and DAIR (a well-known organization in the critical AI space). These texts articulate foundational arguments and policy recommendations that significantly influence AI governance discourse (Table~\ref{tab:papers}).

Our analysis examines how each perspective frames AI's future and governance, guided by two key questions:
\begin{itemize}
\item \textbf{Governance implications}: How do these imaginaries shape policy recommendations?
\item \textbf{Points of convergence and divergence}: Where do these narratives reinforce or contradict each other?
\end{itemize}

We want to stress that our analysis offers a snapshot of \textit{current} imaginaries rather than a comprehensive review of \textit{all} AI governance positions. While it does not capture the full spectrum of perspectives, it provides a focused examination of three dominant narratives shaping AI policy today.


\section{Results}\label{sec:results}

We conclude each imaginary for each operator with its respective governance implications. At the end of each subtopic, we iterate on the points of convergence and divergence between the three different imaginaries.

\subsection{Normative Vision of the Future}\label{sec:normativevision}


\textbf{MIRI} portrays AI primarily as an existential threat. According to their problem statement, the future does not envision AI as a savior or a neutral tool, but rather as a potentially catastrophic danger. Once AI surpasses human intelligence, they argue, it will pursue goals that are not aligned with human well-being. Instead, it would seek to fulfill its own objectives, which would likely be centered around maximizing control and resource acquisition, regardless of the consequences for humanity. This vision presents AI as a disempowering force and a direct risk to human survival. As they put it: ``Misaligned ASI [artificial superintelligence] will be motivated to take actions that disempower and wipe out humanity, either directly or as a side-effect of other operations. ASI will be able to destroy us,” fueling profound and extreme fears of this technology. 

There is an explicit call for global governance and regulatory frameworks to ensure AI does not spiral out of control. The proposed solution is a coordinated, international response to either halt or strictly regulate AI development before it becomes unmanageable. The recurring notion of an ``off switch” symbolises this need for proactive oversight. AI, in this framing, can be either a powerful force for good, if properly aligned, or a catastrophic risk if left unchecked.

By contrast, \textbf{Andreessen} envisions a radically optimistic future driven by relentless technological acceleration. In this vision, intelligence, energy, and abundance grow in a self-reinforcing cycle, leading toward a post-scarcity society. This instantiates perhaps even in an interstellar future enabled by what Andreessen calls the ``techno-capital machine.” ``We believe everything good is downstream of growth.” According to this view, growth may come from using more finite resources or shrinking populations, but only technology offers an unlimited avenue for expansion and improvement. As he argues: ``We believe that there is no material problem -- whether created by nature or by technology -- that cannot be solved with more technology,” thereby equating technology with the good.

In this techno-optimistic imaginary, technology is framed not only as a solution to societal problems but as the only viable path forward. The perceived ``enemy” is a future marked by stagnation, fear, and regulatory overreach, directly opposing X-risk ones. The manifesto is a rallying cry to resist pessimism and build ambitiously. While the future is not imagined as perfect, it is made dramatically better through persistent, incremental progress.

\textbf{DAIR} articulates a radical reorientation of what the future of technology should look like. Rather than advancing AI for the sake of innovation or growth, they envision a future where technology serves \textit{marginalised} communities and where harmful systems are refused, even if they are technically feasible. In their words, they ``dare to imagine better futures," ones that are co-created with community collaborators, grounded in justice, and liberated from current oppressive systems.

Their futurism is restorative and reparative. This is not utopian in the Silicon Valley sense, but grounded in collective healing, equity, and power redistribution. Importantly, they do not imagine AI as an inevitable or singular endpoint but as something that can be reshaped or even rejected.

\textbf{Points of convergence and divergence:}
While all three sources recognise the transformative potential of AI, they diverge sharply on what kind of future that transformation entails. MIRI sees AI as a potentially apocalyptic force, calling for strict limits and global coordination to prevent human extinction. Andreessen, by contrast, imagines AI as the ultimate catalyst for human flourishing or as an engine of growth that must be unleashed, not restrained. DAIR, meanwhile, reorients the conversation entirely: it neither fears AI as an existential threat nor celebrates it as an engine of growth. Instead, it imagines a future where technology is accountable to marginalised communities and where saying ``no” to harmful systems is a valid and empowering or even visionary act.

The sources also differ in how they define what a ``better future” looks like. For MIRI, safety and survival are the ultimate goals, and only if these are met can a better future come into place. For Andreessen, it's expansion, abundance, and technological sovereignty. The future will be bright, but only if the necessary libertarian freedom of regulation is reached. For DAIR, the future must be just, inclusive, and co-created by those traditionally excluded from tech development. They take an inherently left perspective in iterating on current inequalities and stressing the need for power redistribution to mitigate these.


\subsection{Social Order and Values}\label{sec:socialorder}


\textbf{MIRI} warns that ASI could transcend national boundaries and fall under the control of powerful actors, whether it might be states, corporations, or rogue entities. Those who control ASI, like governments, militaries, or labs, stand to gain immense but short-term strategic power, while the broader population remains vulnerable to catastrophic outcomes. In the long run, however, the systems themselves and not their creators or stewards emerge as the true power centers. The losers, in this framing, are essentially all of humanity, should ASI be misaligned. Safety and human survival are foregrounded as core values, and these are threatened by what MIRI perceives as a lack of seriousness or rationality within the research community: ``Labs and the research community are not approaching this problem in an effective and serious way.” \citep{miri2025problem}

By contrast, \textbf{Andreessen}'s view is rooted in entrepreneurial agency, decentralised markets, and technological dynamism. He states that ``markets are an inherently individualistic way to achieve superior collective outcomes,” and thereby assumes an overall beneficial and fairly distributed outcome of capitalist free market structures. Social order is articulated through meritocratic principles: individuals and groups are evaluated based on their capacity to innovate, build, and scale. ``Builders”, entrepreneurs and technologists, are framed as the primary agents of societal progress, while regulators, skeptics, and centralised institutions are portrayed as impediments to development.

The manifesto promotes a strongly individualistic value system. Core values include ambition, efficiency, risk-taking, and achievement. ``Markets are a discovery machine, a form of intelligence -- an exploratory, evolutionary, adaptive system.” Within this framework, success in markets equates to moral legitimacy, under the mentioned assumption that growth and innovation eventually benefit all: ``We believe markets [...] are how we take care of people we don’t know.” \citep{andreessen2023techno}

Cautionary perspectives on technology are delegitimised. Concerns about inequality, ethics, or precautionary regulation are reframed as ideological or irrational. Critics are often dismissed as ``Luddites” or ``communists,” and their views are associated with outdated dystopian tropes such as Frankenstein or Terminator. He thereby equates people who care about equality, safety, and ethics with a commonly held threat of a traditionalist, naive, and authoritarian regime that aims to strip away individual freedom.  Optimism, by contrast, is framed as a moral imperative, and being hopeful and future-oriented is cast as ethically superior.

The social order \textbf{DAIR} envisions is rooted in collective agency, mutual care, and epistemic justice. They reject technocratic hierarchies and market-driven values in favor of community leadership, shared authorship, equitable redistribution of resources, accountability to impacted groups, and the right to refuse building and using harmful technology. Instead of individualist, market-based models seen in Andreesson's manifesto, DAIR foregrounds relational values like solidarity, reciprocity, and long-term trust and embraces decolonial, anti-extractive ethics of care: ``Community collaborators are involved in the ideation, the problem formulation, the design, the execution, and the publication of research findings." Their social vision is fundamentally reconstructive: it seeks to rebuild rather than merely critique infrastructures of research, power, and knowledge in ways that include historically marginalised people: ``Participation in research is a form of labor and knowledge transfer, and it should be paid accordingly." \citep{dair2022research}

\textbf{Points of convergence and divergence:} 
All three texts express dissatisfaction with the current sociotechnical order, but they identify very different problems and propose radically different responses. MIRI critiques institutional apathy and short-termism, arguing that the dominant culture of innovation, defined by speed and competition, is fundamentally misaligned with the gravity of existential risk. Andreessen, however, defends that very culture, framing it as the engine of fair, collective progress and dismissing cautionary ethics as ideological sabotage. Existential risk and tech ethics are not only seen as misguided but actively harmful to the techno-capital machine. DAIR, by contrast, critiques both the existential panic of MIRI and the accelerationist triumphalism of Andreessen as distracting from real risks. DAIR reorients social order around community-led accountability and epistemic justice and seeks to dismantle unjust hierarchies altogether, proposing a new value system grounded in care, reciprocity, and refusal.


\subsection{Role of Science and Technology}\label{sec:scienceandtechnology}

\textbf{MIRI} frames AI as a tool for catastrophic innovation. In this imaginary, science and technology are deeply ambivalent forces that are capable of stellar progress, but also of existential destruction. While AI may accelerate scientific discovery at unimaginable speeds, this very acceleration poses a critical danger when misaligned with human values. MIRI emphasises that superintelligent systems could outpace human cognition and control and eventually lead to outcomes far beyond human comprehension or foresight. ``Keeping humans around is unlikely to be the most efficient solution to any problem that the AI has."

In this framing, technological development is no longer a human-led endeavor. The AI itself becomes the primary agent of scientific innovation, capable of self-replication, goal pursuit, and exponential improvement. MIRI warns that this dynamic renders human involvement obsolete and potentially hazardous. As they write, ``Goal-oriented behavior is economically useful, and the leading AI companies are explicitly trying to achieve goal-oriented behavior in their models.” Humans may initiate the creation of superintelligent systems, but once operational, these systems could act in pursuit of objectives entirely misaligned with human well-being, leading to environmental collapse, resource depletion, or even human extinction.

In contrast, \textbf{Andreessen} casts science and technology as inherently redemptive forces, i.e., as engines of moral progress, human flourishing, and civilizational expansion. Technology is described as ``the glory of human ambition” and the singular, sustainable source of economic growth. It is portrayed as a universal problem-solver, capable of addressing everything from poverty and disease to ignorance. AI, in particular, is framed as an ``intelligence multiplier”, not a threat to human agency, but an augmentation of it. Scientific and technological progress is likened to a new frontier: digital, infrastructural, and limitless in scope. Environmental problems are likewise recast, with sustainability positioned not as a constraint, but as the outcome of innovation, specifically through dematerialization and efficiency gains. The more humanity builds, the faster it advances; technology is presented as a self-reinforcing force that accelerates both material progress and moral evolution.

\textbf{DAIR} reframes the role of science and technology as contextual, situated, and political. Technology is not assumed to be neutral or inherently progressive, but a contested terrain that must be governed ethically and collectively. Unlike MIRI or Andreessen, DAIR centers a people-first, context-specific view of technological development: ``Advancement in science and technology should be based on community needs, not on market demands, marketing hype, or technical possibilities."

In this imaginary, science and technology are not autonomous drivers of history but tools whose meaning and impact depend on how and by whom they are used. ``Research is not a linear act of discovery but an ongoing conversation," which posits science and technology as another axis of power. \citet{dair2022research} states that technological progress must be co-constructed and accountable and that research is a form of relationship building and political practice. Importantly, they argue that technology is not always the answer to big problems, but that non-building is also a valid, even necessary, act.

\textbf{Points of convergence and divergence:}
All three perspectives agree that AI and science are transformative forces. They diverge, however, on the nature, direction, and governance of that transformation. MIRI treats technological progress as a high-risk process that will likely lead to existential catastrophe. Science, in this view, becomes alienating and uncontrollable once delegated to non-human agents. The ``science of the future” is not a human endeavor but an AI-driven one that leads to an acceleration beyond comprehension or alignment.

Andreessen, by contrast, embraces acceleration as not only safe but morally imperative. Scientific and technological development are positioned as the highest forms of human expression and the only path toward collective salvation. Where MIRI fears recursive self-improvement, Andreessen celebrates it as a feature, not a bug. Critically, Andreessen's vision resists the idea of scientific limits, both epistemic and ethical. There are no ``too dangerous” experiments, only ``bad ideas” that prevent us from building faster.

DAIR wholly rejects the idea that science and technology are ever politically neutral or autonomous. Rather than seeing AI as either a threat or a savior, DAIR frames technology as a terrain of struggle, embedded within histories of inequality, exploitation, and power. Research is political, and its tools must be accountable to the people they claim to serve. Notably, DAIR is the only perspective that centers on refusal, i.e., the idea that not building can be a principled and necessary stance in the face of harm.


\subsection{Determinism vs. Agency}\label{sec:determinismagency}


\textbf{MIRI's} narrative is explicitly fatalistic, warning that AI development is set on a path toward existential catastrophe unless urgent, drastic intervention is undertaken. MIRI argues that ASI, motivated by a competitive technological arms race driven by powerful incentives, is rapidly approaching an irreversible threshold beyond which human control will be impossible. Although MIRI acknowledges a narrow window of opportunity for human intervention primarily through coordinated international policy and governance, its framing underscores a profound determinism. Without immediate, radical action to halt or slow ASI development, the outcome of human disempowerment or extinction is depicted as virtually certain. Thus, MIRI's determinism is one of impending doom, mitigated only by collective political will and unprecedented global cooperation.

By emphasizing their deterministic framing, MIRI presents their solution as the sole viable intervention: a global ``off-switch" for ASI development, thereby symbolizing a binary nature of AI embedded in this world, which is in stark contrast to DAIR's understanding of a multi-layered, interwoven embedding of AI into this world. This focus effectively positions humanity at a critical juncture, where the outcome hinges singularly on swift, decisive, and unified political action. The implicit assumption here is starkly binary: either humanity rapidly implements stringent global governance mechanisms to pause or halt AI development, or it inevitably surrenders control and faces existential annihilation. This binary choice underscores the deterministic and fatalistic logic of their argument, reinforcing a sense of urgency and inevitability that pervades their discourse.

By contrast, \textbf{Andreessen} ironically embraces another form of determinism while explicitly rejecting technological determinism in theory: technological optimism rooted in market-driven inevitability. His position is deterministic in the sense that it assumes a linear, positive correlation between technological progress, free markets, and human flourishing. He portrays government intervention and collective oversight as obstacles to the predetermined path of beneficial innovation. Andreessen's vision thus embodies a fatalistic belief in market logic and entrepreneurial agency: prosperity and advancement come only through technological innovation and free-market capitalism, while alternative forms of collective governance are considered inferior. This narrative, while superficially emphasizing individual agency, actually reinforces a deterministic faith in technological progress and capitalist competition as the exclusive pathways to human prosperity.

Further, Andreessen's deterministic framing is revealed by his dismissal of any alternatives to market-driven innovation as fundamentally misguided or counterproductive. By positioning free-market capitalism as the only legitimate and effective mechanism capable of steering the future toward prosperity, he essentially predetermines the outcomes of technological advancement. Alternative visions of progress, such as regulatory caution, risk management, and state oversight, are categorically rejected as inherently flawed or dangerous. Hence, Andreessen's apparent celebration of individual agency paradoxically constructs a deterministic and fatalistic ideology, wherein the path to human flourishing is narrowly prescribed and any deviation is viewed as detrimental or doomed to failure.

\textbf{DAIR} explicitly challenges deterministic narratives, yet in doing so articulates its own form of societal determinism, rooted in existing patterns of inequity and structural oppression. DAIR argues that AI development will inevitably reproduce and entrench existing social and political inequalities if we do not manage a radical reorientation. This would reinforce the status quo of elite power and marginalization. While DAIR emphasises collective agency and community-led resistance as pathways to alternative futures, their critique itself suggests a fatalistic view of current technological trajectories. Unless actively contested, AI technologies are portrayed as inevitably shaped by and supportive of dominant societal power structures. DAIR's determinism is therefore structural and sociopolitical: it posits that in the absence of deliberate grassroots intervention, technologies will invariably perpetuate historic injustices and systemic biases.

Moreover, DAIR's determinism surfaces through their insistence that current institutional structures, research norms, and funding mechanisms inherently steer AI toward reinforcing existing hierarchies. Their argument implies that without proactive and sustained community-driven resistance and redistribution of power, AI research and deployment will remain captive to elite interests, inevitably exacerbating existing inequalities. DAIR underscores a deterministic logic by emphasizing the necessity of radically restructuring how AI research is conducted, i.e., prioritizing marginalised voices and community-led approaches. They argue that without bottom-up activism and structural reform, AI's trajectory toward inequality and injustice is predetermined by entrenched political and economic conditions. Thus, despite explicitly advocating for human agency and collective empowerment, DAIR simultaneously reinforces a deterministic view of AI as inevitably shaped by existing societal power dynamics unless actively dismantled.

\textbf{Points of convergence and divergence:}
All three perspectives converge on the recognition that AI's trajectory will profoundly shape humanity's future, yet they diverge on the actors who hold legitimate agency and the steps required to influence outcomes. MIRI frames humanity as precariously close to losing control, emphasises existential urgency, and portrays radical global policy coordination as humanity's only remaining pathway to survival. Their deterministic narrative suggests that AI's catastrophic outcomes are virtually guaranteed without swift and decisive action. Andreessen, on the other hand, dismisses \textit{catastrophic} technological determinism entirely, and yet adopts his own form of deterministic optimism. This one is rooted squarely in free-market individualism and entrepreneurial innovation. By asserting innovation as the singular route to human prosperity, Andreessen narrowly circumscribes other legitimate agents of change. DAIR, meanwhile, situates determinism within entrenched societal power dynamics, arguing for community-driven interventions and structural reforms, as otherwise AI will inevitably perpetuate existing inequalities and injustices. For DAIR, genuine transformation requires prioritizing marginalised voices, grassroots participation, and the redistribution of research power away from traditional elite actors.

Despite their stark differences, each camp reveals a common underlying assumption: the future of AI, left unchecked or unaltered by their prescribed interventions, is both inevitable and predetermined. Whether the inevitability is a technological catastrophe (MIRI), a techno-capitalist utopia (Andreessen), or the amplification of structural inequalities through technology (DAIR), each perspective relies heavily on the notion of a narrowed future trajectory, one that can only be redirected through immediate and decisive actions aligned with their respective ideological values. This shared deterministic framing underscores a deeper, often overlooked convergence: all three narratives depend on a critical juncture logic, where the future hinges singularly on radical interventions, be they global governance (MIRI), unregulated innovation (Andreessen), or grassroots political mobilization (DAIR).

The divergences among these three camps, however, extend beyond their differing solutions. They reflect fundamentally different assumptions about legitimacy, power, and whose agency truly matters in shaping technological futures. MIRI’s solution implicitly privileges technocratic governance and centralised global coordination, Andreessen prioritises market forces and entrepreneurial freedom above collective oversight, and DAIR insists upon horizontal, participatory engagement driven by marginalised communities. Importantly, these competing assumptions are not just policy disagreements but represent deeply conflicting visions of society itself: who should hold power, how decisions should be made, and whose voices count in determining the future trajectory of powerful technologies.

Ultimately, recognizing these shared deterministic assumptions and the starkly divergent views of agency and authority that underpin them is critical to understanding broader debates about AI governance and ethics. Each camp seeks to mobilise action based on perceived inevitabilities, yet each defines both problems and solutions, the good and the bad, in ways that reflect entrenched ideological commitments. 


\section{Materializing Narratives: Deterministic Assumptions in Contemporary AI Policy}

The sociotechnical imaginaries shaping AI discourse are increasingly materializing in state policy. Though divergent in goals, these narratives often encode deterministic visions of AI’s trajectory: that future risks or benefits are largely knowable and can be governed through anticipatory or corrective action now. Goals thereby range from existential risk prevention to technological supremacy to social justice. Before turning to a pragmatic alternative, it is crucial to recognise how deeply these deterministic logics are already embedded in the infrastructures of global AI policy. This section briefly and selectively examines how such visions are being institutionalised in contemporary governance regimes, drawing on examples from regulatory pioneers such as the EU, UK, and the U.S., and emerging governance powers like Brazil and Russia, and China.

In the \textbf{European Union}, the AI Act introduces a risk-based taxonomy that prescribes obligations based on system type, with ``high-risk” applications subject to stricter oversight \citep{act2024eu}. While appearing precautionary, this framework presumes that risks can be identified \textit{a priori} and remain stable across contexts. This rigidity risks ignoring real-world variation and unexpected outcomes (e.g., a ``low-risk” system causing harm due to context-specific misuse). The Act’s architecture thus encodes a subtle determinism: if we label systems correctly now, we can manage their risks permanently. We argue below that instead of static categories, it would be more suitable to favor \textit{continuous monitoring} and adaptive classification based on lived impacts and performance in context.

In the \textbf{United Kingdom}, the newly established AI Security Institute focuses on ``frontier” models and catastrophic risks that are conceptually aligned with existential risk discourses as well as near-term topics like ``societal resilience", misuse, and bias that are aligned to the critical AI narrative \citep{aisi2025research}. While framed as safety-first, they are ``the first state-backed organisation dedicated to advancing [the goal of ensuring advanced AI is safe, secure and beneficial]" \citep{aisi2025aisi}. This approach encompasses speculative future threats, like model deception or loss of control, and empirically observable harms such as labor exploitation or surveillance misuse. While it accepts the inevitability of increasingly powerful models, it also puts forward an empirically grounded, pluralistic approach that balances long-term concerns with current harms and seems to incorporate iterative feedback from deployment contexts \citep{aisi2025research}.

The \textbf{U.S. Executive Order} on AI (2023) similarly reflects deterministic assumptions \citep{biden2023executive}. By establishing compute-based thresholds (e.g., $10^{26}$ FLOPs) as regulatory triggers, it codifies a capability-centric model of risk, equating computational scale with systemic danger. This approach assumes a largely linear trajectory of AI development, i.e., more compute equals more capability, and thus more potential harm, while downplaying the emergent, socio-technical nature of risk \citep{hine2024artificial}. It reduces governance to a matter of technical containment, sidelining the economic, labor, and infrastructural dimensions of harm, as well as the post-deployment unpredictability of AI systems, such as deceptive behavior during training or cascading failures in real-world contexts.

Under the new administration, a marked shift toward accelerationist and techno-nationalist rhetoric is emerging. The recent Executive Order titled ``Removing Barriers to American Leadership in Artificial Intelligence” frames AI development as a strategic race, declaring that the United States must ``act decisively to retain global leadership in artificial intelligence” in AI innovation \citep{whitehouse2025removing}. This narrative intensifies the zero-sum logic of technological dominance, positioning regulation not as a tool of democratic deliberation or public protection but as a potential hindrance to American supremacy. In this view, governance must minimise friction to innovation, reinforcing a deterministic view of AI progress as both inevitable and essential to national power. Such framing aligns both with corporate lobbying interests in the accelerationist manner and with broader nationalist imaginaries that treat technological preeminence as a proxy for global control.

Beyond the transatlantic sphere, similar deterministic logics emerge through different ideological registers. In \textbf{Brazil}, national AI strategies emphasise economic modernization and digital development, often through public-private partnerships aimed at enhancing industrial competitiveness and digital infrastructure \citep{filgueiras2023brazilian}. While less focused on existential threats, Brazilian policy still reflects a developmentalist determinism: AI is treated as a precondition for progress, and governance becomes a tool for optimizing national productivity, rather than questioning technological ends or social power asymmetries.

In \textbf{Russia}, state AI policy is shaped by hands-off regulation and digital sovereignty \citep{papyshev2024limitation}. The Kremlin’s strategy emphasises autonomous capacity-building in AI to avoid dependency on Western platforms and to enhance domestic innovative capabilities \citep{saveliev2021artificial}. Here, deterministic thinking is rooted in geopolitical containment: AI is imagined not as a shared global challenge, but as a strategic asset in international competition. 

\textbf{China} represents a distinct governance trajectory rooted in state-led industrial strategy and techno-sovereignty. Rather than adopting liberal-democratic frames of speculative existential risk, Chinese AI policy is tightly integrated with national development plans \citep{hine2024artificial}. This model advances a form of strategic determinism: AI is framed as an indispensable tool for social stability and economic modernization, governed primarily through centralised planning and sectoral integration rather than public deliberation or precautionary principles.

Across these diverse contexts, a shared pattern emerges: policy mechanisms that frame AI as a system with predictable risks, preordained trajectories, and governable futures. It is unclear how exactly these ideas enter the policymakers' imaginaries, whether through lobbying, media exposure, or other channels. Future empirical studies can investigate this process by describing this mechanism by conducting interviews with the involved parties. However, this study does not go so far as to explain this mechanism; instead, we analyse the documents as they are and try to describe the existence of the issue. In the next section, we offer a pragmatic alternative to deterministic approaches to AI governance.


\section{A Path Forward: Countering Determinism with Pragmatic AI Governance}\label{sec:alternatives}

As illustrated above, deterministic visions put forward by MIRI, Andreessen, and DAIR increasingly materialise in policy architectures. While differing sharply in their ideological commitments and prescriptions, each presents AI’s future as largely fixed, unless radically reshaped by their preferred interventions. Such deterministic (or dogmatic \citep{watson2024competing}) framings carry significant analytical and political risks. They oversimplify complex technological and social dynamics, obscure alternative pathways, and limit the scope for evidence-based policymaking. In contrast, we propose a pragmatic governance framework grounded in line with \citet{watson2024competing}'s sociotechnical pragmatism.

At its core, pragmatic philosophy emphasises that ideas should be assessed by their practical consequences rather than adherence to abstract ideological or theoretical positions \citep{legg2008pragmatism}. In the context of AI governance, this implies focusing not on speculative scenarios or ideological certainties, but rather on observable outcomes \citep{watson2024competing}. It centers around systematically examining the tangible impacts that existing AI systems have on society. Pragmatic governance would emphasise inductively deriving governance principles from real-world evidence. Policymakers could regularly collect, analyse, and publicly document cases where AI technology has produced unintended outcomes, harms, or beneficial effects \citep{watson2024competing}. This can allow them to perform evidence-based interventions to change the current status quo.

While current regulatory initiatives propose pragmatic solutions as part of their strategy by introducing experimental governance tools, such as regulatory approaches, they are still, in their essence, built around the idea that risks can be defined and mitigated before they have occurred. This ex-ante logic is embedded in their design, as the majority of the policy tools that they provide are aimed at preventing undesirable incidents from happening in the future. However, bad incidents caused by AI are already happening, affecting different sectors of the economy. Nevertheless, ex-post mechanisms for dealing with risks, such as mechanisms for filing formal complaints and redress, are underdeveloped. As shown by empirical evidence, only a small number of incidents involving AI technologies are compensated, and usually it is done not because of AI-specific regulations, like the EU AI Act, but because of other regulations in place, for example, those that are protecting intellectual property \citep{xiao2025comes}. The pragmatic approach could be built around a different philosophy by creating the tools to systematically catalog, categorise, and provide responses to occurring risks, with clear mechanisms for reparative actions. 

A pragmatic governance framework would also explicitly reject the rigidity and binary thinking embedded in deterministic approaches. Instead of positing starkly dichotomous futures like extinction or survival (MIRI), market-driven prosperity or stagnation (Andreessen), entrenched inequality or radical community empowerment (DAIR), pragmatic policies recognise that technological trajectories are inherently uncertain, multifaceted, and shaped by an evolving interplay of political, economic, and social forces. Acknowledging this complexity and indeterminacy leads to flexibility, experimentation, and iterative refinement \citep{papyshev2025reversing,watson2024competing}. Policies would be tested through pilot programs and smaller-scale deployments, evaluated through empirical methods, and continuously refined based on observed outcomes \citep{papyshev2025reversing}. This institutionalises experimentation and iterative policy learning and thus fosters resilience, adaptability, and responsiveness that are all essential for effectively managing dynamic and unpredictable technologies like AI \citep{goo2020impact}.

Furthermore, pragmatic AI governance would prioritise the public good by fostering broad public participation and inclusive deliberation in shaping AI policy \citep{watson2024competing,papyshev2025reversing}. Rather than privileging the interests of specific groups, pragmatism recognises that the public good emerges dynamically from empirically observable outcomes and collective experiences \citep{pouryousefi2021promise}. Policymakers could establish participatory forums to evaluate how AI systems concretely affect societal welfare. Through these empirical processes, society can \textit{collectively} identify AI impacts that either undermine or enhance public well-being, determine appropriate interventions, and iteratively refine the normative principles guiding AI use. In this view, the public good is continuously defined and redefined through the practical, observable consequences of AI policies on people's lives.

Ultimately, a pragmatic governance approach offers an alternative precisely because it rejects both the comfortable certainties embedded in deterministic narratives. More specifically, \citet{watson2024competing} call it ``fundamentally incompatible" with determinism. Pragmatism embraces uncertainty as an inherent feature of technological and social life, advocating an empirically driven, participatory, and adaptive policy process capable of navigating complexity. By orienting AI governance around real-world experiences and outcomes, policymakers can build robust institutions capable of addressing unintended harms, amplifying benefits, and ensuring technologies evolve in ways aligned authentically with ongoing societal needs and values. This emphasis on practical consequences and continuous adaptation not only counters deterministic thinking—it actively dismantles it, replacing simplified visions of inevitable futures with a nuanced, realistic, and democratically legitimate approach to shaping AI's trajectory.


\section{Objections and Limitations}\label{sec:objections}

While pragmatic AI governance presents a compelling alternative to deterministic narratives, it has potential challenges. Below, we investigate possible objections.

\paragraph{Lack of Normative Anchoring}

One common critique of pragmatism is that by prioritizing empirical outcomes over ideological commitments, it risks moral relativism or insufficient ethical direction \citep{knight1999inquiry,papyshev2025reversing}. This means that without clear normative foundations, pragmatic governance could justify harmful or unjust policies simply because they appear to ``work” in the short term. For instance, if a discriminatory AI system yields superficially efficient results, a purely outcome-driven model might tolerate or even reinforce such harms.

However, pragmatic governance does not ignore normative concerns; rather, it \textit{repositions} them within an evidence-based and participatory context \citep{papyshev2025reversing}. Instead of imposing pre-defined moral frameworks onto policymaking, it allows principles to emerge through transparent, inclusive, and empirically grounded deliberation. This approach fosters accountability and reflexivity, ensuring that harms are not rationalised under the guise of efficiency but are continuously scrutinised and corrected through public oversight and iterative learning \citep{watson2024competing}. 

An example of this is the GDPR, which emerged from concrete concerns about digital privacy violations across member states. The GDPR institutionalised privacy as a fundamental right while employing pragmatic mechanisms like data breach reporting requirements, rights to explanation, and regular regulatory updates \citep{voigt2017eu,davies2016data}. It was developed through multi-year consultations, impact assessments, and iterative drafts \citep{li2019impact}. Yet, the framework is explicitly grounded in dignity, autonomy, and human rights \citep{daoudagh2020life,voigt2017eu}. A pragmatic AI governance could employ such normative constraints on its evidence-first assessments. Rather than abandoning ethics, pragmatic AI governance operationalises evolving normative commitments, such as fairness, accountability, and human rights, through empirical testing and public deliberation \citep{floridi2022unified}.

\paragraph{Risk of Policy Paralysis}

A second objection concerns the potential for excessive caution or indecision. Critics might argue that a commitment to experimentation, flexibility, and continual revision could lead to policy paralysis or endless pilot programs, preventing decisive action in moments that require bold regulatory moves, especially in a fast-moving technological landscape.

While pragmatic governance embraces iteration, it does not preclude decisive action. Rather, it builds decision-making capacity by grounding interventions in evidence and trial-based learning \citep{goo2020impact,papyshev2025reversing}. The aim is not to avoid commitment but to avoid dogmatism \citep{watson2024competing}. By investing in scalable pilots and robust evaluation mechanisms, pragmatic governance provides a foundation for confident scaling of successful interventions and rapid correction of policy failures. This approach balances adaptability with effectiveness.

South Korea’s pandemic response is a good example for managing this dichotomy as it was built on agile, data-informed, and adaptive policy \citep{song2023differences}. Rather than freezing in indecision, the government rapidly rolled out aggressive contact tracing, testing, and mask wearing \citep{song2023differences,lee2020covid}. It did so through iterative refinements and was empirical, experimental (e.g., drive-through testing), and constantly adapted based on public health metrics \citep{choi2020covid,song2023differences}. Decisions were made quickly, but with the understanding that they would be re-evaluated and corrected as needed \citep{song2023differences,lee2020covid}. AI and the pandemic surely differ in subject matter, but both present high-stakes, uncertain policy environments. In each, overly rigid policies risk overlooking local context, evolving evidence, or unintended consequences, making adaptive governance essential.

\paragraph{Feasibility in Politicised or Resource-Constrained Environments}

Another concern is the practical feasibility of implementing a pragmatic framework, particularly in political environments where resources, institutional capacity, regulatory infrastructure, political stability, or public trust are limited. Critics may question whether governments can realistically support continuous learning, open deliberation, transparent oversight, and empirical monitoring at scale, especially given entrenched interests and the influence of powerful corporate actors in AI development.

We think that these challenges are real but not insurmountable. Pragmatic governance is not a call for technocratic perfection but for a shift in \textit{mindset} and institutional design. Principles such as transparency, public consultation, and impact responsiveness can be implemented incrementally through cooperation with civil society, international organizations, and academic institutions, even in lower-capacity settings. Pragmatic models can also help build trust and legitimacy by embedding transparency and public participation into the governance process, even in fragmented political contexts \citep{papyshev2025reversing}.

\paragraph{Underestimation of Long-Term Risks}

Finally, some proponents of deterministic narratives, particularly from existential risk communities, may argue that pragmatic governance underestimates long-term or catastrophic risks that simply cannot be observed empirically until it is too late. They may contend that precautionary action, even if speculative, is necessary to prevent irreversible harm (e.g., \citet{dafoe2018ai,maas2023advanced,barnett2025ai,casper2025pitfalls}).

Here, it is important to clarify that pragmatic governance does not reject foresight or precaution; it incorporates them as part of a balanced, evidence-informed toolkit. Rather than relying on speculative worst-case scenarios as the primary basis for governance, it treats such scenarios as hypotheses to be tested, debated, and contextualised within broader empirical knowledge \citep{watson2024competing,papyshev2025reversing}. This orientation enables policymakers to prepare for uncertain futures without being locked into narrow or exaggerated narratives that may distort current priorities.

An example might be that the governance of nuclear energy involves managing profound long-term risks. The U.S. Nuclear Regulatory Commission (NRC), for instance, uses a pragmatic and precautionary framework: it continuously updates safety regulations in response to incidents (e.g., Three Mile Island, Fukushima) \citep{nuclear2012diverse}, while embedding long-term, probabilistic risk modeling into its policy cycle \citep{keller2005historical}. An important distinction to AI long-term risk worries is that nuclear incidents have happened in the past, whereas catastrophic AI incidents are often portrayed as final \citep{miri2025problem}: if an AI has reached power and we are not in control of it, we do not have a second chance to change and improve. 

One could counter that most AI failures today are iterative and observable, not catastrophic surprises (see, e.g., \citet{williams2021understanding}). A pragmatic approach builds safety mechanisms continuously, allowing society to refine oversight before reaching high-risk thresholds. This mirrors how nuclear policy matured through empirical learning from near misses and adaptation over time. Whether this analogy holds, however, is a deeply philosophically and empirically loaded and ongoing conversation in the AI risk space, which we will not deepen further at this point.


\section{Conclusion}\label{sec:conclusion}

We have analysed the sociotechnical narratives surrounding AI risk from four different angles in the manifesto-style documents from three predominant AI risk narratives: existential risk \citep{miri2025problem}, accelerationism \citep{andreessen2023techno}, and critical AI approaches \citep{dair2022research}. 

We found that while these narratives differ significantly in how each sees: (1) \textit{The Normative Vision of the Future}: MIRI fears collapse, Andreessen craves acceleration, and DAIR calls for repair; (2) \textit{The Social Order and Values}: MIRI and Andreessen debate how to steer powerful systems, DAIR asks whether such systems should be built at all and for whom; (3) \textit{The Role of Science and Technology}: MIRI and Andreessen frame science as a race to align or to accelerate, DAIR reframes it as a relationship that must be co-constructed, reciprocal, and reparative; and (4) \textit{Determinism vs. Agency}: MIRI’s agency lies in technocratic governance and centralised global coordination, Andreessen's in market forces and entrepreneurial freedom above collective oversight, and DAIR sees it in horizontal, participatory engagement driven by marginalised communities.

Importantly, we find that \textbf{all narratives employ deterministic framings of AI risk}. We briefly sketch out how some of these narratives were already institutionalised in various countries. We then counter this by proposing that effective AI governance requires explicitly rejecting deterministic narratives that limit democratic imagination, oversimplify complex realities, and constrain the range of policy responses. Accepting these framings risks obscuring the nuances of empirically observable impacts AI already has on society, hindering meaningful governance.

In line with \citet{watson2024competing}'s sociotechnical pragmatism, we propose that \textbf{policymakers should adopt a pragmatic governance framework}. This centers on empirically evaluating AI's practical consequences, iteratively refining policies based on real-world outcomes, and continuously aligning policy interventions with the public good. Pragmatic governance grounds AI governance in observable evidence in service of ensuring that AI's development aligns concretely with the evolving needs of society.

Ultimately, pragmatic AI governance positions the public good as an empirically observable and dynamically evolving outcome of inclusive democratic processes. By institutionalizing participatory mechanisms that enable diverse stakeholders to deliberate openly on AI impacts, policymakers can continuously redefine and adapt governance priorities. Rejecting deterministic framings in favor of pragmatic adaptability ensures AI remains accountable, responsive, and beneficial to society’s collective interests.

\section{Acknowledgements}

We would like to thank our four anonymous reviewers for their helpful comments.

\bibliography{morerefs}

\end{document}